\documentclass[proceedings]{JHEPhepph} % 10pt is ignored!

\usepackage{epsfig}                   % please use epsfig for figures
\usepackage{psfig}                   % please use epsfig for figures
\usepackage{axodraw}                   % please use epsfig for figures

\def\gsim{\:\raisebox{-0.5ex}{$\stackrel{\textstyle>}{\sim}$}\:}
\def\eq#1{{Eq. (\ref{#1})}}

\def\np#1#2#3{           {\it Nucl. Phys. }{\bf #1} (19#2) #3}

\def\n.c.#1#2#3{         {\it Nuovo Cim. }{\bf #1} (19#2) #3}
\def\r.n.c.#1#2#3{       {\it Riv. del Nuovo Cim. }{\bf #1} (19#2) #3}

\def\be{\begin{equation}}       
\def\ee{\end{equation}}
\def\bear{\be\begin{array}}
\def\eear{\end{array}\ee}
\def\bea{\begin{eqnarray}}
\def\eea{\end{eqnarray}}

\def\21{$SU(2) \ot U(1)$}
\def\ot{\otimes}
\def\ie{{\it i.e.}}
\def\etal{{\it et al.}}

\def\bold#1{\setbox0=\hbox{$#1$}
     \kern-.025em\copy0\kern-\wd0
     \kern.05em\copy0\kern-\wd0
     \kern-.025em\raise.0433em\box0 }

\conference{hep-ph/9907220}

\title{The decay {$\bold{b\rightarrow s\gamma}$}  and the
charged Higgs boson mass  without   R-Parity}

\author{E. Torrente-Lujan\thanks{Based on work made in collaboration with M. A. D\'\i az and J. W. F. Valle and presented at: ``TMR  Network on Physics beyond the SM''. Trieste, February 1999.}
\\
      
{ Dept. de F\'\i sica Te\'orica, IFIC-CSIC, 
Universitat de Val\`encia} 
{ Burjassot, Val\`encia 46100, Spain}\\
        E-mail: \email{e.torrente@cern.ch}}

\keywords{Charged Higgs bounds. BRpV. FCNC. SUSY }

\abstract{

The experimental measurement of $B( b \to s\gamma)$ imposes important
constraints on the charged Higgs boson mass in the MSSM.  
We show that by adding bilinear R--Parity violation (BRpV)
in the tau sector, these bounds are relaxed. The bound on
$m_{H^{\pm}}$ in the MSSM--BRpV model is $\gsim 200-250$ GeV for
the the heavy squark limit.
For lighter
squarks, light charged Higgs bosons can be reconciled with $B(b
\to s\gamma)$ only if there is also a light chargino. In the BRpV model if
we impose $m_{\chi^{\pm}_1}>90$ GeV 
 $m_{H^{\pm}} \gsim 75$ GeV, around 30 GeV down from the MSSM. In this case the charged Higgs bosons would be
observable at LEP II. The relaxation of the bounds is due mainly to
the fact that charged Higgs bosons mix with staus and they contribute
importantly to $B(b \to s\gamma)$.}

\begin{document}

\section{Introduction}

In this talk, I will try to show you how the existing limits on the mass of the charged Higgs obtained from the measurement of the  $B(b\rightarrow s\gamma)$ decay by the 
CLEO group relax in a supersymmetric model with bilinear
violation of R-Parity, in the so called $\epsilon$-model, a BRpV model 
 where only the
tau neutrino becomes massive at the tree level
(see Ref.\cite{DTV} for a complete exposition).

All previous work on $b \to s\gamma$ in supersymmetry has assumed the
conservation of R--Parity. 
In the model assumed here new particles 
contribute in the loops to $B(b\rightarrow s\gamma)$. Charginos mix
with the tau lepton 
(this mixing is not in conflict with the well measured tau couplings to 
gauge bosons \cite{ADV}),  
therefore, the tau lepton contribute to the decay rate together with 
up-type squarks in the loops. Nevertheless, this 
 contribution can be neglected \cite{DTV}. In a similar way, the 
charged Higgs boson mixes with the two staus \cite{ChaHiggs} forming a 
set of four charged scalars, one of them being the charged Goldstone 
boson. In this way, the staus contribute to the decay rate together with 
up-type quarks in the loops (see Table \ref{tab1}). 

\begin{center}
\begin{table}
\scalebox{0.6}{
\begin{tabular}{ccc}
\begin{picture}(120,90)(0,21) % y_2 controlates equation position
\ArrowLine(20,25)(60,25)
\Vertex(60,25){2}
\PhotonArc(85,25)(25,0,180){2}{8}
\ArrowLine(60,25)(110,25)
\Vertex(110,25){2}
\ArrowLine(110,25)(150,25)
\Photon(100,5)(70,15){1}{6}
%\ArrowLine(60,25)(110,55)
\Text(85,34)[]{$u_k$}
\Text(85,61)[]{$W^-$}
\Text(35,34)[]{$b$}
\Text(135,34)[]{$s$}
\Text(85,-5)[]{$\gamma,Z^0,g$}
\end{picture}
    &\hspace{1cm} &
\begin{picture}(120,90)(0,21) % y_2 controlates equation position
\ArrowLine(20,25)(60,25)
\Vertex(60,25){2}
\DashCArc(85,25)(25,0,180){5}
\ArrowLine(60,25)(110,25)
\Vertex(110,25){2}
\ArrowLine(110,25)(150,25)
\Photon(100,5)(70,15){1}{6}
%\ArrowLine(60,25)(110,55)
\Text(85,34)[]{${u}_k,t,\tau$}
\Text(85,61)[]{$S_j^-$}
\Text(35,34)[]{$b$}
\Text(135,34)[]{$s$}
\end{picture}
\\
\begin{picture}(120,90)(0,21) % y_2 controlates equation position
\ArrowLine(20,25)(60,25)
\Vertex(60,25){2}
\ArrowArcn(85,25)(25,180,0)
\DashLine(60,25)(110,25){5}
\Vertex(110,25){2}
\ArrowLine(110,25)(150,25)
\Photon(100,5)(70,15){1}{6}
%\ArrowLine(60,25)(110,55)
\Text(85,34)[]{$\tilde{u}_k,\tilde{t},\tilde{\tau}$}
\Text(85,61)[]{$\tilde{F}_j^-$}
\Text(35,34)[]{$b$}
\Text(135,34)[]{$s$}
\end{picture}
&\hspace{1cm} &
\begin{picture}(120,90)(0,21) % y_2 controlates equation position
\ArrowLine(20,25)(60,25)
\Vertex(60,25){2}
\ArrowArcn(85,25)(25,180,0)
\DashLine(60,25)(110,25){5}
\Vertex(110,25){2}
\ArrowLine(110,25)(150,25)
\Photon(100,5)(70,15){1}{6}
%\ArrowLine(60,25)(110,55)
\Text(85,34)[]{$\tilde{d}_k$}
\Text(85,61)[]{$N=\tilde{\chi}_j^0, \tilde{g},\nu_\tau$}
\Text(35,34)[]{$b$}
\Text(135,34)[]{$s$}
\end{picture}
\end{tabular}
}
\vspace{0.1cm}
\caption{$b\to s\gamma$ penguin diagrams in BRpV. $S$ fields are mixtures
 or charged Higgs and staus. $F$ fields are mixtures or charginos and the
tau lepton.}
\label{tab1}
\end{table}
\end{center}
\section{ FCNC processes and BRpV.}
Gauge invariance,renormalizability and particle content
of the SM imply the absence of FCNC in the lepton sector. 
FCNC transitions in the quark sector are absent at the tree level
(see Ref.\cite{gabbiani,MPR,okada} and references therein for a complete 
review of FCNC processes).
At one loop they are suppressed by light quark masses relative to 
$m_W$ and by small mixing between the third and the first 
generations. The predicted SM suppression of FCNC process is in 
beautiful agreement with the presently available experimental data.

Being rare processes mediated by loop diagrams, 
radiative decays of B mesons are potentially sensitive probes of new physics beyond the SM.
In the context of SUSY models we confront ourselves
 with a generic 
flavor problem. 
The low scale of new physics together with the 
absence of any constraint on the structure of SUSY breaking 
can produce easily huge, disastrous rates for 
the FCNC and LFV transitions.

In the Minimal and Not-so-Minimal Supersymmetric extensions
of the Standard Model (SSM) there are, broadly speaking, two kinds of new contributions to the FCNC transitions:
\begin{itemize}
\item 
Flavor mixing in the 
sfermion, squark and slepton, mass matrices.
FCNC processes as $b\to s\gamma$ and others  may depend on 
the structure of these matrices  and its 
experimental observation could provide some 
insight on the SUSY breaking mechanism.

\item Charged Higgs boson and chargino ex\-chan\-ges.

\end{itemize}

The dangerous contributions coming from s\-fer\-mion mass ma\-trices 
can be suppresses supposing s\-fer\-mions  of the two 
first generations: generic but very heavy or, 
alternatively de\-gene\-rate in mass.
But, from the second point, there will always remain a minimal flavor violation coming from
 the KM angles present  in vertices 
as  charged Higgs-top and  chargino-squark loop contributions.
FCNC processes, in this case, 
could provide some insight on the structure of the, possibly 
 enlarged, charged boson and fermion sectors.

Charged Higgs-top and chargino-squark vertices are two kind 
 of vertices modified  by the introduction of Bilinear R-parity
violation, so it seems relevant   to study 
the  $b\to s\gamma$ decay in the framework of a BRpV model and  
natural 
to consider 
a minimal flavor violation 
hypothesis as it  will be 
done  on this work: 
we will suppose  
that all squarks other than the scalar partners of the top quark 
have the same mass $\tilde{m}$ and that   contributions from 
s\-fer\-mion mass matrices can be neglected completely. 
In practice, that 
means that we will neglect contributions coming from gluino and 
neutralino loops. 

\section{ The CLEO Experimental Results.}
In 1995, the CLEO collaboration reported the first measurement of the 
inclusive branching ratio for the radiative decays 
$B\to X_s\gamma$. 
This measurement has established for the
first time the existence of one--loop penguin diagrams. 
The latest presented result is \cite{CLEO}:
$$ B(B\to X_s\gamma)= (3.15\pm 0.35\pm 0.32\pm 0.26)\times 10^{-4}. $$
Form here, one obtains an upper limit  $BR<4.5$ at the 95\% CL (note that the bound is one sided).

In addition, the ALEPH collaboration has reported \cite{CLEO} a measurement for the corresponding 
branching ratio for b hadrons produced at the 
Z resonance, yielding 
$ B(H_b\to X_s\gamma)= (3.11\pm 0.80\pm 0.72)\times 10^{-4}. $
Theoretically the two numbers are expected to differ by at most a few 
per cent, the weighted average gives:
$ B(B\to X_s\gamma)= (3.14\pm 0.48)\times 10^{-4}. $

In order to reject photon background only the 
high energy part of the photon produced in the b decay 
is accessible experimentally. 
This value quoted above is  obtained by extrapolation to the 
low energy part of the photon spectrum and is 
model dependent \cite{neubert}.
This fact 
introduces a significant theoretical 
uncertainty ($\approx 7\%$ in the last CLEO measurement).

QCD corrections are very important and can be a substantial fraction
of the decay rate. Recently, several groups have completed the
Next--to--Leading order QCD corrections to $B(b \to s \gamma)$.
Two--loop corrections to matrix elements were calculated in
\cite{GHW}. The two--loop boundary conditions were obtained in
\cite{AY}. Bremss\-trah\-lung corrections were 
obtained in \cite{bremss}. Finally, three--loop anomalous dimensions
in the effective theory used for resumation of large lo\-ga\-rithms
 were found in \cite{CMisiakM,anomalous}). In this work we include all these QCD corrections.
The present 
 theoretical incertitude  is slightly less than the experimental 
 errors: $<\approx 10\%$.
During the years 1996-1997 the next-to-leading order analysis was extended to 
the cases of two-Higgs doublet models and MSSM \cite{bor1}.

In the SM, loops including the $W$ gauge boson and the unphysical
charged Goldstone boson $G^{\pm}$ contribute to the decay rate. 
The up-to-date SM theoretical value is ($E_{\gamma}> 0.1 E_\gamma^{max}$):
\begin{eqnarray}
Br(B\to X_s\gamma)&=&(3.29\pm 0.33)\times 10^{-4} N_{LO},\nonumber
\end{eqnarray}
Where $N_{LO}= Br(B\to X_c e\overline{\nu})/0.108\simeq 1$.
This prediction is in
agreement with the CLEO measurement at the $2\sigma$ level.
It expected some improvements in the   experimental error
 coming from the possibility of a better  rejection of background and the measurement of the emitted photon spectrum.
However, possible new physics contributions would not affect 
the shape of this photon spectrum. They  would enter the 
theoretical predictions for 
the $BR(B\to X_s\gamma)$ through the values of 
the Wilson coefficients at the scale $M_W$.

\section{The $\epsilon$-BRpV model}
The study of models which include BRpV terms, and not trilinear (TRpV), 
is motivated by spontaneous 
R--Parity breaking \cite{spont,SRpSB2}. 
The superpotential we consider here contains the following 
bilinear terms which violates
R--Parity and tau--lepton number explicitly. 
\begin{equation} 
W_{Bi}=\varepsilon_{ab}\left[
-\mu\widehat H_1^a\widehat H_2^b
+\epsilon_3\widehat L_3^a\widehat H_2^b\right]\,,
\nonumber
\end{equation}
where both parameters $\mu$ and $\epsilon_3$ have units of mass.

The electroweak symmetry is broken when the two Higgs 
doublets $H_{1,1}$ and the third component of the left slepton 
doublet acquire non-zero Vev's by
the presence of the extra term 
(respectively $v_{1,2}$ and $v_3$).
We define  the angles $\beta$ and $\theta$
in spherical coordinates
$v_1 = v\cos\beta\sin\theta$,
$v_2 = v\sin\beta\sin\theta$,
$v_3 = v\cos\theta$
where $v=246$ GeV and the MSSM relation $\tan\beta=v_2/v_1$ is preserved \cite{e3others,e3others1}. 
The $v_3$ is related to the
mass parameter $\epsilon_3$ through a minimization condition. This
non--zero sneutrino vev is present even in a basis where the
$\epsilon_3$ term disappears from the superpotential.  This basis is
defined by the rotation $\mu'\widehat H_1'=\mu\widehat
H_1-\epsilon_3\widehat L_3$ and $\mu'\widehat L_3'=\epsilon_3\widehat
H_1+\mu\widehat L_3$, with
 $\mu'^2=\mu^2+\epsilon_3^2$.
The sneutrino vev in this basis ($v'_3$) is
non--zero due to mixing terms that appear in the soft sector between
$\widetilde L_3$ and $H_1$ scalars.
It is also possible to choose a basis where the sneutrino vev is
zero. In this basis a non--zero $\epsilon_3$ term is present in the
superpotential \cite{BKMO}. All three basis are equivalent.

The BRpV model has the attractive 
feature of generating masses radiatively for the two first generations, thus naturally small in this framework. 
The origin of the tau neutrino mass is 
 linked to supersymmetry 
\cite{SUSYorig}
through the mixing of neutral higgsinos and gauginos with the neutrino.
In a see--saw type of mechanism, with the neutralino masses 
($\sim$ M) playing the
role of a high scale and $v'_3$ as the low scale, the tau--neutrino mass
is approximately given by the expression
(for  $m_{\nu_\tau}$ small and $M$ not so small):
\begin{displaymath}
m_{\nu_{\tau}}\simeq -({g^2+g'^2})\ v'^2_3/4 M.
\end{displaymath}
In addition, the combination  $v_3'$
is radiatively induced in soft universal models and then
  naturally small.
The dependence of the tau--neutrino 
mass on  $v'_3$  can be appreciated in Fig .~\ref{mneutrino} where 
results of a random scan are shown (see below). 
We easily find solutions with neutrino masses from the collider limit
of 20 MeV down to eV. 
The experimental bound on the tau neutrino
mass, given by $m_{\nu_{\tau}}<18$ MeV \cite{tauBound}, implies an
upper bound for $v'_3$ of about 5--10 GeV. 

%%% exec b003.exec goto plotneutrino
\FIGURE{
\protect\hbox{\psfig{file=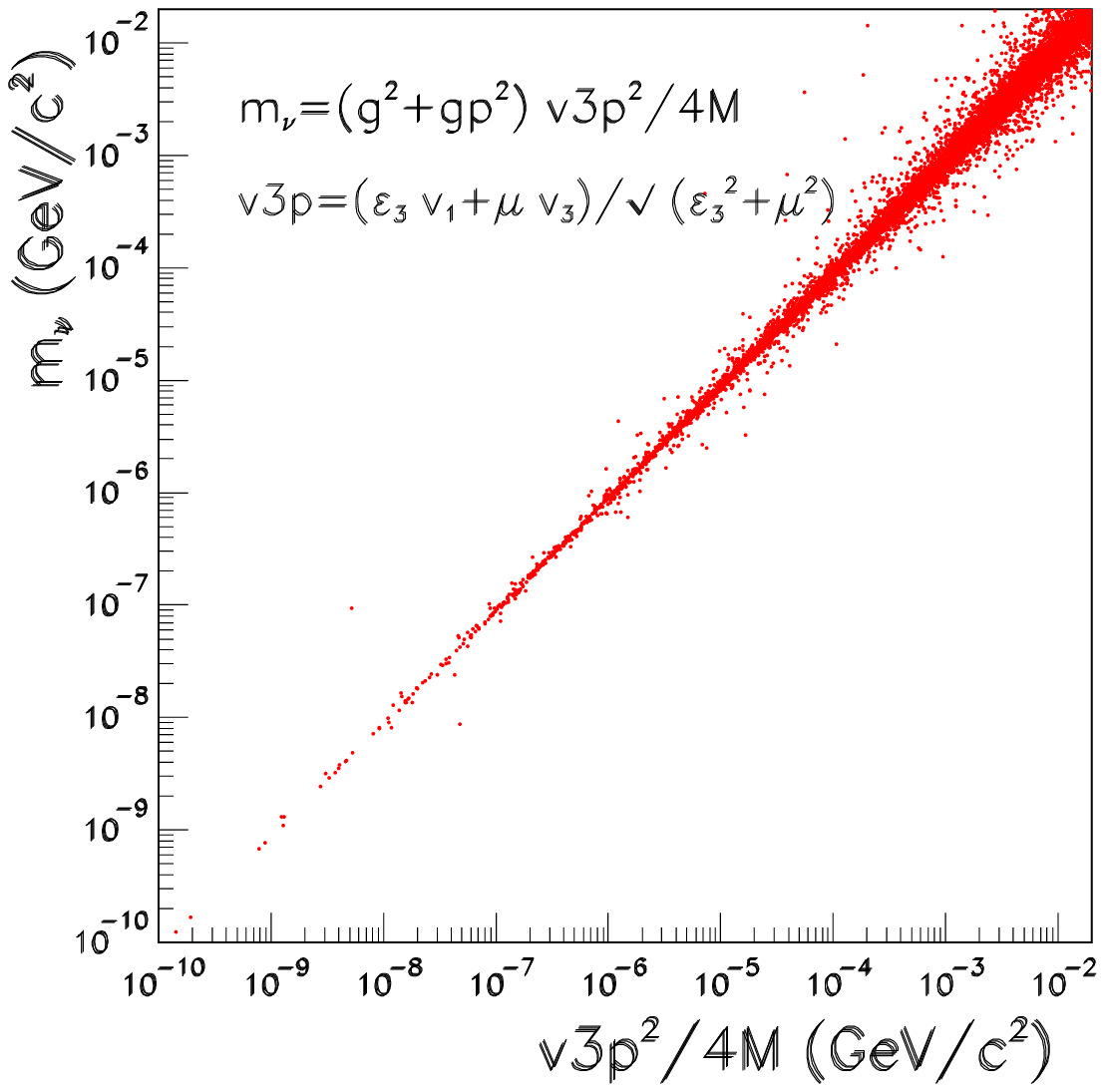,height=4.5cm}}
\caption{Correlation of $m_\tau$ 
 with $v_3'$ for the points belonging to the MSSM-BRpV  
parameter  space scan (see text). }
\label{mneutrino}
}

\section{ $H^\pm$ mass and $b\to s\gamma$ decay}
Direct 
search limits on charged Higgs bosons are provided by LEP.
However, by far the most restrictive process in constraining the charged Higgs sector in 2HDM is  the radiative b-quark decay, 
These constraints are specially important in 2HDM type II models
because the charged Higgs contribution always adds to the SM
contribution \cite{HewettBBP}.
Constraints on $m_{H^{\pm}}$ are not important in 2HDM type I because
charged Higgs contributions can have either sign.

In supersymmetric models, loops containing 
char\-ginos/squarks, 
neu\-tra\-li\-nos/squarks, and glu\-ino/squarks have to be included \cite{BBMR}.
In the limit of very heavy super-partners, the stringent bounds on 
$m_{H^{\pm}}$ are valid in the MSSM, a 2HDM-II model \cite{HewettBBP}. 
Nevertheless, even in this case the bound is relaxed at large $\tan\beta$
due to two-loop effects \cite{bsgDiaz1}. It was shown also that by 
decreasing the squarks and chargino masses this bound disappears because 
the chargino contribution can be large and can have the opposite sign to 
the charged Higgs contribution, canceling it \cite{SUSY1,BSG}. 
Further studies have been made in the MSSM and in its Supergravity version 
\cite{bsgLater,bsgRecent}. As a result, for example, most of the parameter 
space in MSSM-SUGRA is ruled out for $\mu<0$ especially for large 
$\tan\beta$ (\cite{BBMR,SUSY1} see also Ref.\cite{DTV} and references therein).

Relative to the calculation of $B(b \to s\gamma)$, the main difference
of MSSM-BRpV with respect to the MSSM is that in BRpV the charged
Higgs boson mixes with the staus and the tau lepton mixes with the
charginos.  This way, new contributions have to be added and the old
contributions are modified by mixing angles.

 In the MSSM--BRpV, the
charged Higgs sector mixes with the stau sector forming a set of four
charged scalars. 
The four charged scalars in the original basis are ${\bf\Phi}^{\pm}=
(H_1^{\pm},H_2^{\pm},\tilde\tau_L^{\pm},\tilde\tau_R^{\pm})$ and the
corresponding mass matrix is dia\-go\-na\-li\-zed after the rotation
${\bf S}^{\pm}={\bf R}_{S^{\pm}}{\bf\Phi}^{\pm}$ where ${\bf S}^{\pm}_i$,
$i=1,2,3,4$ are the mass eigenstates (one of them the unphysical Goldstone 
boson). One of the massive charged scalars has similar properties to
the charged Higgs of the MSSM. In BRpV we call the ``charged Higgs boson''
to the charged scalar whose couplings to quarks are larger, \ie, 
maximum $({\bf R}_{S^{\pm}}^{i1})^2+({\bf R}_{S^{\pm}}^{i2})^2$.
Nevertheless, for comparison we have also study the case in which the
``charged Higgs boson'' corresponds to the charged scalar with largest 
components to the rotated Higgs fields $H'^{\pm}_1$ and $H'^{\pm}_2$, 
\ie, maximum $({\bf R'}_{S^{\pm}}^{i1})^2+({\bf R'}_{S^{\pm}}^{i2})^2$.

 In BRpV, the tau lepton mixes with the char\-ginos forming a
set of three char\-ged fer\-mions $F_i^{\pm}$, $i=1,2,3$. In the
original basis where $\psi^{+T}= (-i\lambda^+,\widetilde
H_2^1,\tau_R^+)$ and $\psi^{-T}= (-i\lambda^-,\widetilde
H_1^2,\tau_L^-)$, the charged fer\-mion mass 3x3 matrix 
 $\bold M_C$ is of the form:
\begin{equation}
{\bold M_C}=\left[\matrix{
M & {\textstyle{1\over{\sqrt{2}}}}gv_2 & 0 \cr
{\textstyle{1\over{\sqrt{2}}}}gv_1 & \mu & 
-{\textstyle{1\over{\sqrt{2}}}}h_{\tau}v_3 \cr
{\textstyle{1\over{\sqrt{2}}}}gv_3 & -\epsilon_3 &
{\textstyle{1\over{\sqrt{2}}}}h_{\tau}v_1}\right]
\nonumber
\end{equation}
where   $\tau$ Yukawa coupling $h_\tau$ is  a complicated
 function of SUSY parameters and is fixed by the condition
$m_\tau=1.77$ GeV. In the not-so-small $M$ limit we
recover a simple expression for $h_\tau$ 
($v_1'\equiv (\mu v_1 -\epsilon_3 v_3)\mu'$):
$$h_\tau\simeq\surd 2 m_\tau/v_1'.$$
 In the $b\to s\gamma$  amplitude 
appears the Wilson coefficients $C_{7,8}$ at the 
$\sim M_w$ scale (see Ref.\cite{DTV} for concrete expressions):
\begin{displaymath}
C_{7,8}(M_w)\sim A_{\gamma,g}= A^W_{\gamma,g}
+A_{\gamma,g}^{F^{\pm}}
+A_{\gamma,g}^{S^{\pm}}
+A_{\gamma,g}^{\chi^{0},\tilde{g}}
\end{displaymath}
where $A^W$ is the SM contribution and $A^{F,S}$ the contributions
of the $F,S$ fields defined previously. The running down the
$\sim m_b$ scale has been performed using the corrections
developed in Refs.\cite{CMisiakM,MPR}.
According to our hypothesis of minimal flavor violation, we neglect
in our calculations neutralino and gluino 
($A_{\gamma,g}^{\chi^{0},\tilde{g}}$)
contributions. In any case the contribution of neutralinos 
is small \cite{BBMR} as it is  small that one  of the gluino whose different squark 
contributions tend to cancel with each other \cite{BaerEtal}. In addition, 
if gaugino masses are universal at the GUT scale, gluinos must be rather 
heavy 
 considering the bound on the chargino mass from LEP2 \cite{DiazKing},
which makes the contribution smaller. We can ignore  safely 
the light gluino window 
\cite{Clavelli} because it is inconsistent with the experimental bound on
the mass of the lightest Higgs boson in the MSSM \cite{Diazgluino}.
In the decay amplitude \cite{DTV}, it appears
the matrix $R_{\tilde t}$  the rotation matrix which
dia\-go\-na\-li\-zes the stop quark mass matrix \cite{epsrad} necessary to
take into account the left--right mixing in the stop mass matrix. We
neglect this mixing for the other up--type squarks.

\section{Results. The parameter space scan}
In order to study the effect of BRpV on $B(b \to s\gamma)$ we consider
 the so-called unconstrained MSSM--BRpV  where  all soft 
parameters are independent at the weak scale, \ie, not embedded into
supergravity. We study the predictions of this model 
 varying randomly the soft parameters
at the weak scale \cite{DTV}.
The scan over parameter space  contains over $5\times 10^4$
points in the  ranges:
\begin{eqnarray}
&\mid \mu,B\mid          &<500 \,\,{\mathrm{GeV}}\,,
\nonumber\\
 0.5<&\tan\beta      &<30 \,,
\nonumber\\
  10<&M_{L_3},M_{R_3}&<1000 \,\,{\mathrm{GeV}}\,,
\nonumber\\
 100<&M_Q=M_U        &<1500 \,\,{\mathrm{GeV}}\,,
\nonumber\\
  50<&M = 2M'          &<1000 \,\,{\mathrm{GeV}}\,,
\nonumber\\
&\mid A_t,A_{\tau}\mid   &<500 \,\,{\mathrm{GeV}}
\label{paramMSSM}
\end{eqnarray}
for the MSSM parameters, and
\begin{eqnarray}
&\mid\epsilon_3\mid &<200 \,\,{\mathrm{GeV}}\,,
\nonumber\\
&\mid v'_3   \mid    &<10 \,\,{\mathrm{GeV}}
\label{paramBRpV}
\end{eqnarray}
for the BRpV parameters.

In addition, in order to study 
more in detail the neutrino low mass region 
 we have  performed a dedicated 
scanning  for masses 
$m_\nu <100 \,\,{\mathrm{eV}}$.

In \eq{paramMSSM}, $B$ is the bilinear soft
mass parameter associated with the $\mu$ term in the superpotential,
$M_{L_3}$ and $M_{R_3}$ are the soft mass parameters in the stau
sector, $M_Q$ and $M_U$ are the soft mass parameters in the stop
sector. The parameters $A_t$ and $A_{\tau}$ are the trilinear soft
masses in the stop and stau sector respectively. Note that $B_2$, the
bilinear soft mass parameter associated with the $\epsilon_3$ term in
the superpotential, is fixed by the minimization equations of the
scalar potential.

\subsection{Discussion}

In order to have an idea of the effects of BRpV on the constraints
from the measurement of $B(b \to s\gamma)$ it is instructive to take
the limit of very massive squarks. In this limit the chargino
amplitude  can be neglected relative to the charged
scalar amplitude and a lower limit
on the MSSM charged Higgs mass is inferred.

In Fig.~\ref{brmh1} we
plot the branching ratio $B(b \to s\gamma)$ as a function of the
charged Higgs mass $m_{H^{\pm}}$ in the MSSM with large squark masses
(in practice, masses at least equal to several TeV are necessary to
suppress the chargino amplitude). The horizontal dashed line
corresponds to the latest CLEO upper limit.
In Fig.~\ref{brmh2} we plot the $B(b \to s\gamma)$ as a function of
$m_{H^{\pm}}$ in the MSSM--BRpV model in the same 
heavy squark limit. The
difference is exclusively due to the mixing of the charged Higgs boson
with the staus.  
The clear bound we had before dissapears.
The reason for this relaxation  is
simple. We have now new contributions in the 
charged boson sector and 
while the charged Higgs couplings to quarks diminish due to
Higgs--Stau mixing, the contribution from the staus does not always
compensate it, because staus may be heavier than the charged Higgs
boson.

\FIGURE{
\protect\hbox{\psfig{file=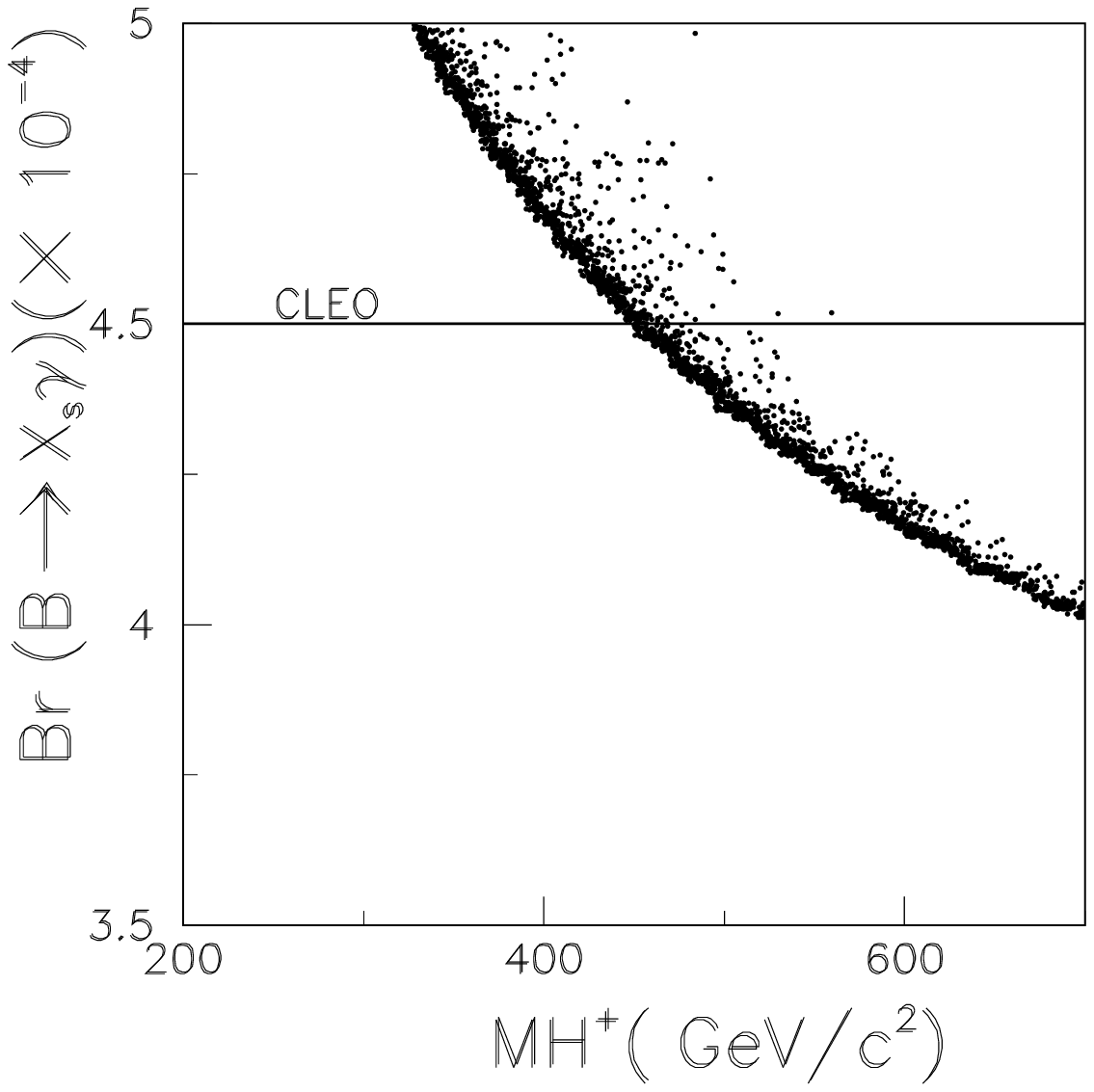,height=4cm}}
\caption{Branching ratio $B(b \to  s\gamma)$ as a function of the 
charged Higgs boson mass $m_{H^{\pm}}$ in the limit of very heavy
squark masses within the MSSM.}
\label{brmh1}
}
\FIGURE{
\protect\hbox{\psfig{file=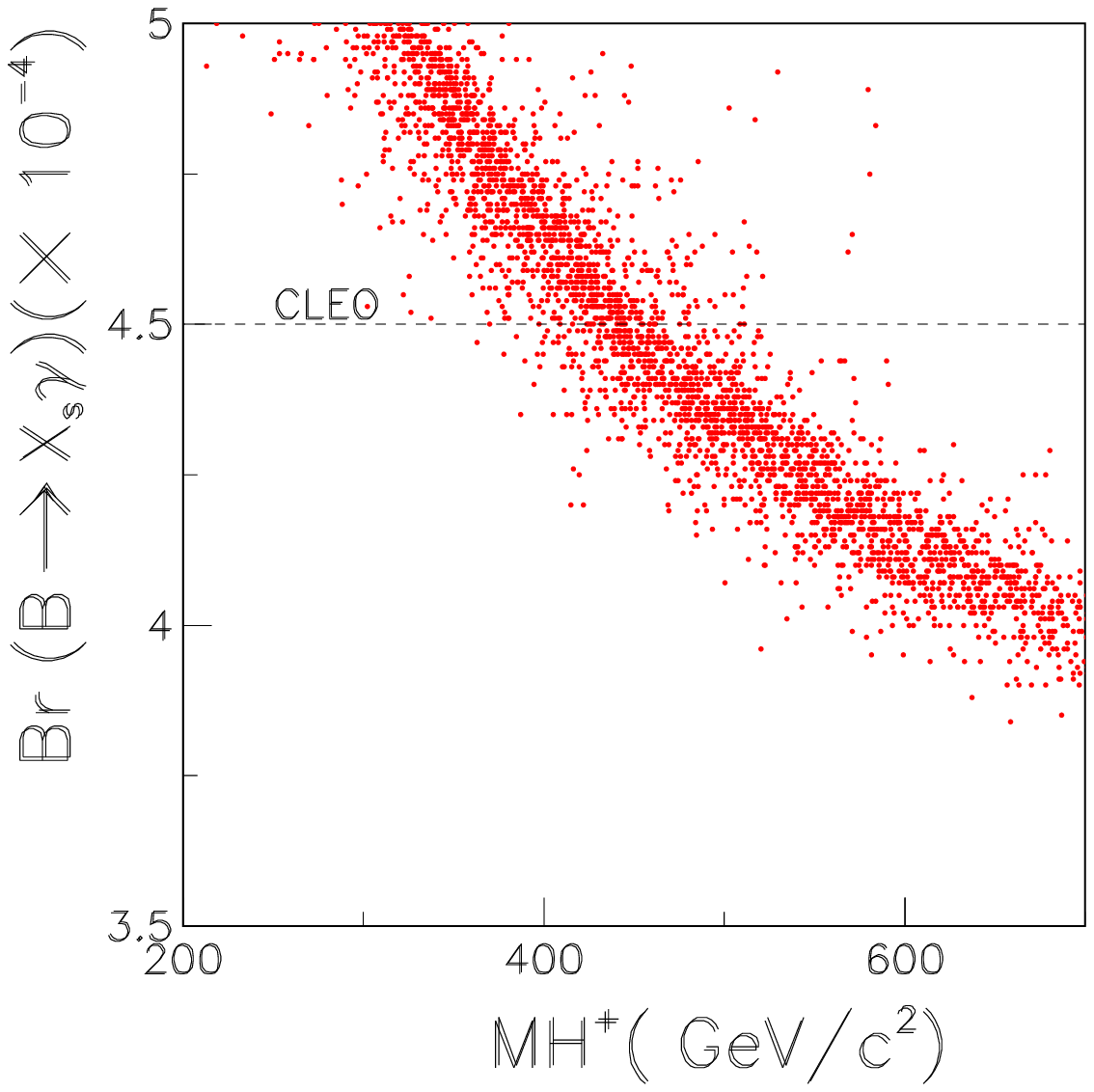,height=3.8cm}}
\caption{Branching ratio $B(b \to  s\gamma)$ as a function of the 
charged Higgs boson mass $m_{H^{\pm}}$ in the limit of very heavy
squark masses in MSSM--BRpV. The charged Higgs boson is defined as the 
massive charged scalar field with largest couplings to quarks.}
\label{brmh2}
}
A summary of the results  can be
appreciated in Fig.~\ref{brmh}. 
In the limit of very heavy squarks, 
the strong constraints imposed on the 
charged Higgs mass of the MSSM are relaxed in the MSSM--BRpV. Above and to the right of the solid line in the figure 
are the solutions of the MSSM consistent with the CLEO measurement of 
$B(b\rightarrow s\gamma)$. 
Without considering theoretical uncertainties, 
the limit on the charged Higgs mass is $m_{H^{\pm}}>440$ GeV. This bound 
is relaxed by about 70 to 100 GeV in BRpV as can be seen from the dotted 
and dashed lines. If a 10\% theoretical uncertainty is considered, the 
MSSM bound reduces to $m_{H^{\pm}}>320$ GeV, but the BRpV bound 
decreased 
as well such that the reduction of the bound is maintained. 
\FIGURE{
\protect\hbox{\psfig{file=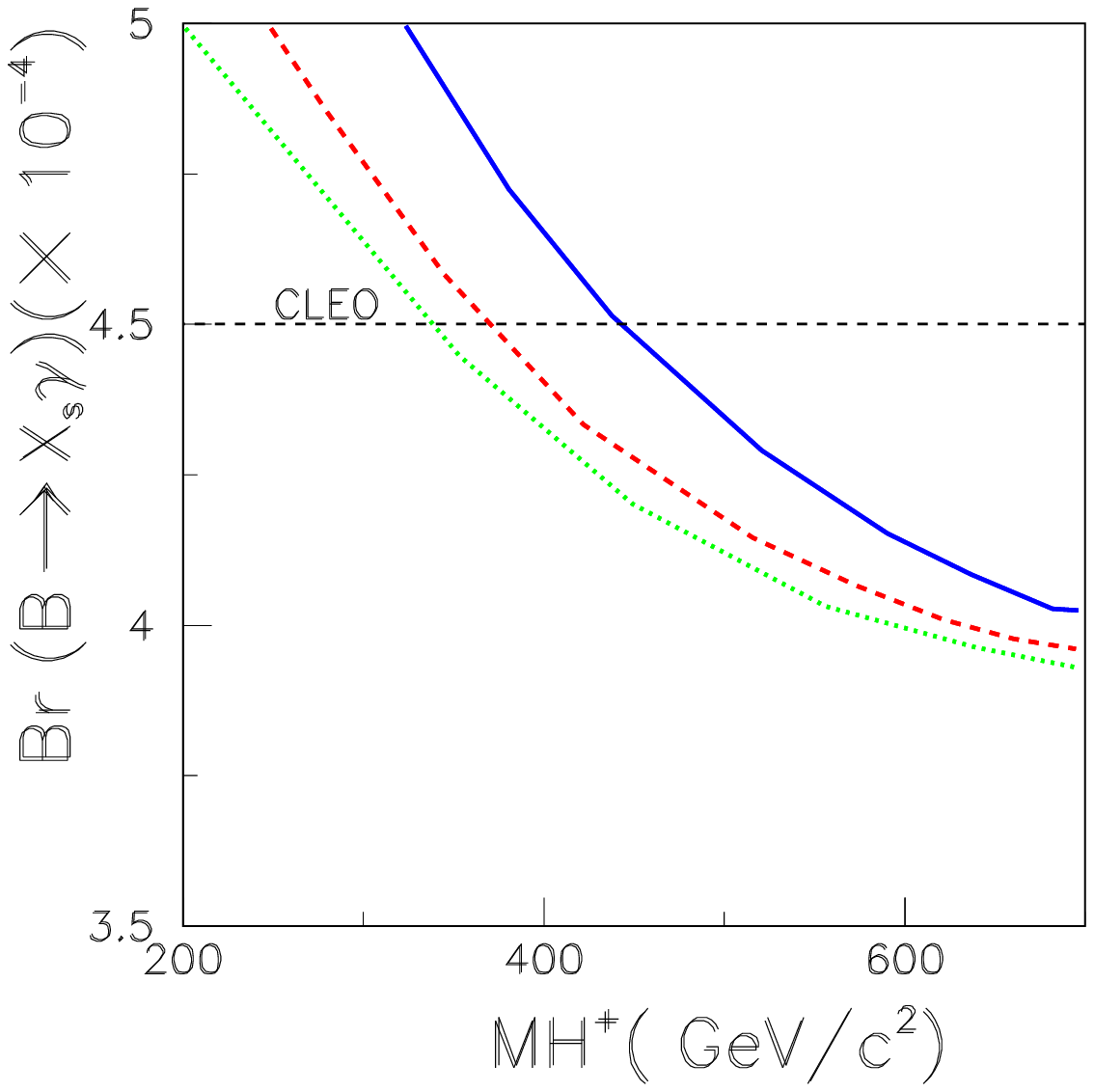,height=4.cm}}
\caption{Lower limit on the branching ratio $B(b \to  s\gamma)$ as 
a function of the charged Higgs boson mass $m_{H^{\pm}}$. We consider 
the limit of very heavy squark masses within the MSSM (solid) and 
the MSSM--BRpV (dashes and dots).}
\label{brmh}
}

Another interesting region of parameter space to explore is the region of
light charged Higgs boson and light chargino. It is known that in order 
to have a light charged Higgs boson, its large contribution to 
$B(b \to  s\gamma)$ must be canceled by the contribution from
light charginos and stops. 

In Fig.~\ref{mhcha} we give the lower bounds on $m_{H^{\pm}}$
as a function of the lightest chargino mass $m_{\chi^{\pm}_1}$. 
All the points satisfy the CLEO
bound mentioned before. The solid vertical line is defined by 
$m_{\chi^{\pm}_1}=90$ GeV, which is approximately the experimental
lower limit found by LEP2, at least for the heavy sneutrino case.
The 
solid curve corresponds to the MSSM limit 
and the dotted curve corresponds to the MSSM--BRpV limit.
From the figure, we  observe
that in order to have $m_{\chi^{\pm}_1}>90$ GeV, 
the CLEO measurement of $B(b \to  s\gamma)$ implies that
$m_{H^{\pm}}\gsim 110$ GeV in the MSSM. 
However, in the MSSM-BRpV, in order to have $m_{\chi^{\pm}_1}>90$ GeV 
compatible with $B(b \to  s\gamma)$ we need $m_{H^{\pm}}\gsim 85$ 
GeV, therefore, relaxing the MSSM bound by about 25 GeV. 
\FIGURE{
\protect\hbox{\psfig{file=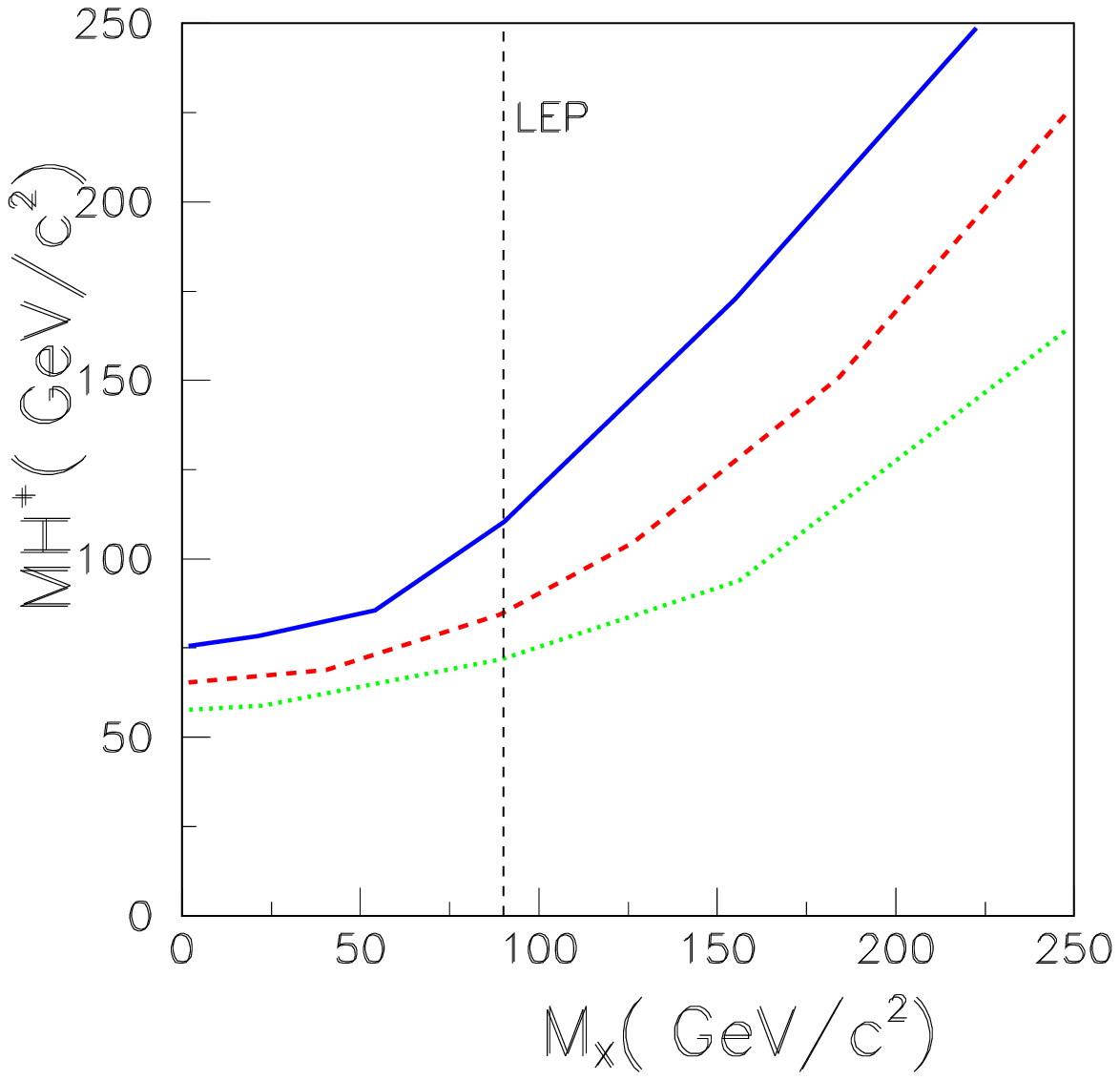,height=4.cm}}
\caption{Lower limit of the charged Higgs boson mass as a function of
the lightest chargino mass for $B(b \to s\gamma)$ compatible with CLEO
measurement in the MSSM (solid) and in MSSM--BRpV (dashes and dots as
explained in the text).  The vertical dashed line corresponds to
$m_{\chi_1}=90$ GeV.}
\label{mhcha}
}

In the same way,
in Fig.~\ref{mhstp} we plot the same lower bounds on $m_{H^{\pm}}$ but 
this time as a function of the lightest stop mass $m_{\tilde t_1}$. We 
observe from this figure that in order to cancel large contributions
to $B(b \to  s\gamma)$ due to a light charged Higgs boson, it is
more important to have a light chargino rather than a light stop.

\FIGURE{
\protect\hbox{\psfig{file=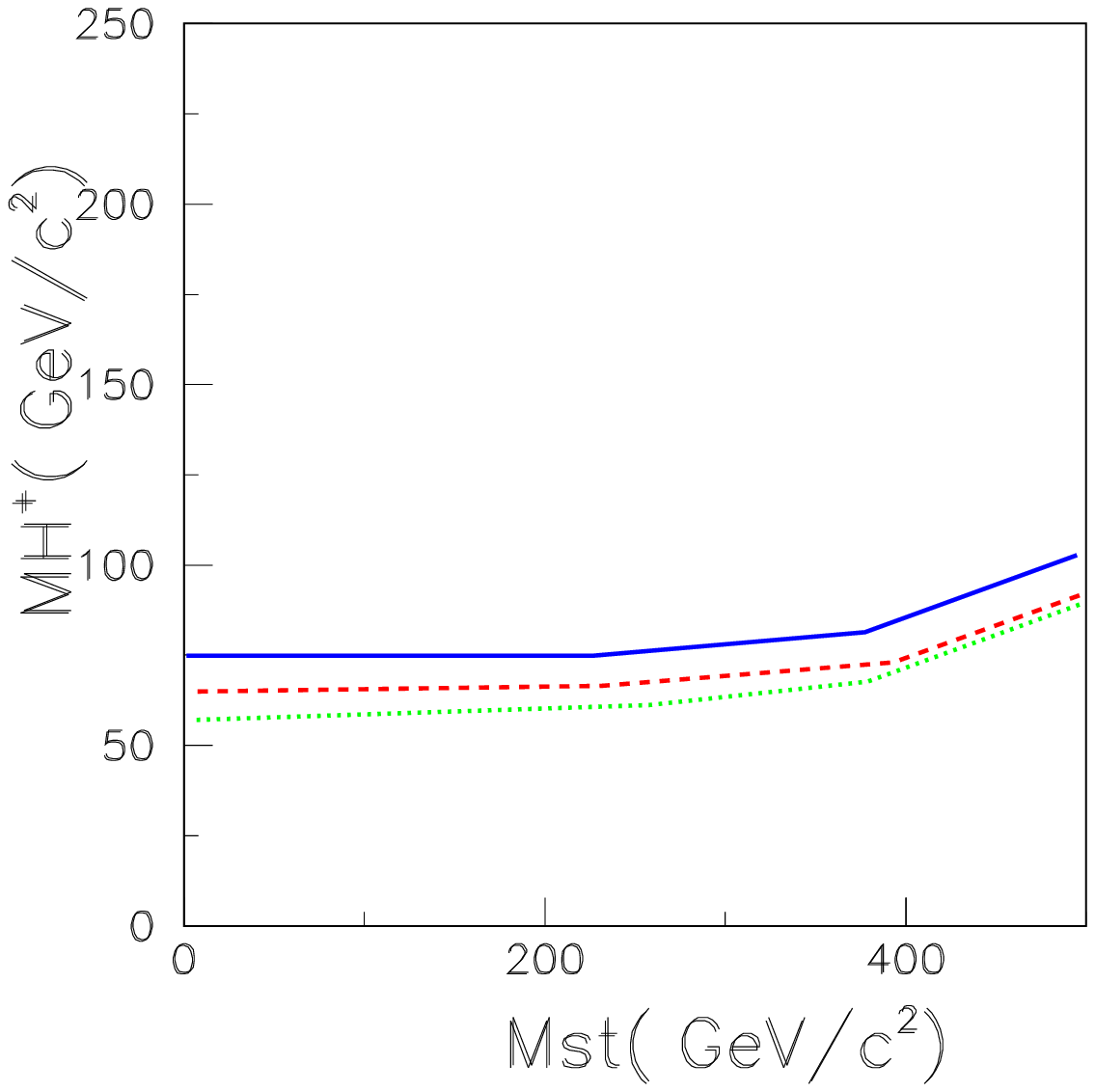,height=3.5cm}}
\caption{Lower limit of the charged Higgs boson mass as a function of
the lightest stop mass for $B(b \to s\gamma)$ compatible with CLEO
measurement in the MSSM (solid) and in the MSSM--BRpV (dashes and dots
as explained in the text).}
\label{mhstp}
}

An important point to see is what happens for low neutrino masses.
We have checked explicitly that our results remain if we impose 
progressively stronger cuts on the neutrino mass.
As example, in Fig.(\ref{lowneutrino}) we show what happens
when the neutrino mass is down $100$ eV in both,  the heavy 
squark and the light chargino limits. We see that there is no
significant differences with respect to the plots we have 
shown before. It is difficult to say anything
 definitive much below the $100 $ eV region because  radiative 
corrections could play an important role at very low neutrino masses.
\FIGURE{
\begin{tabular}{ll}
\protect\hbox{\psfig{file=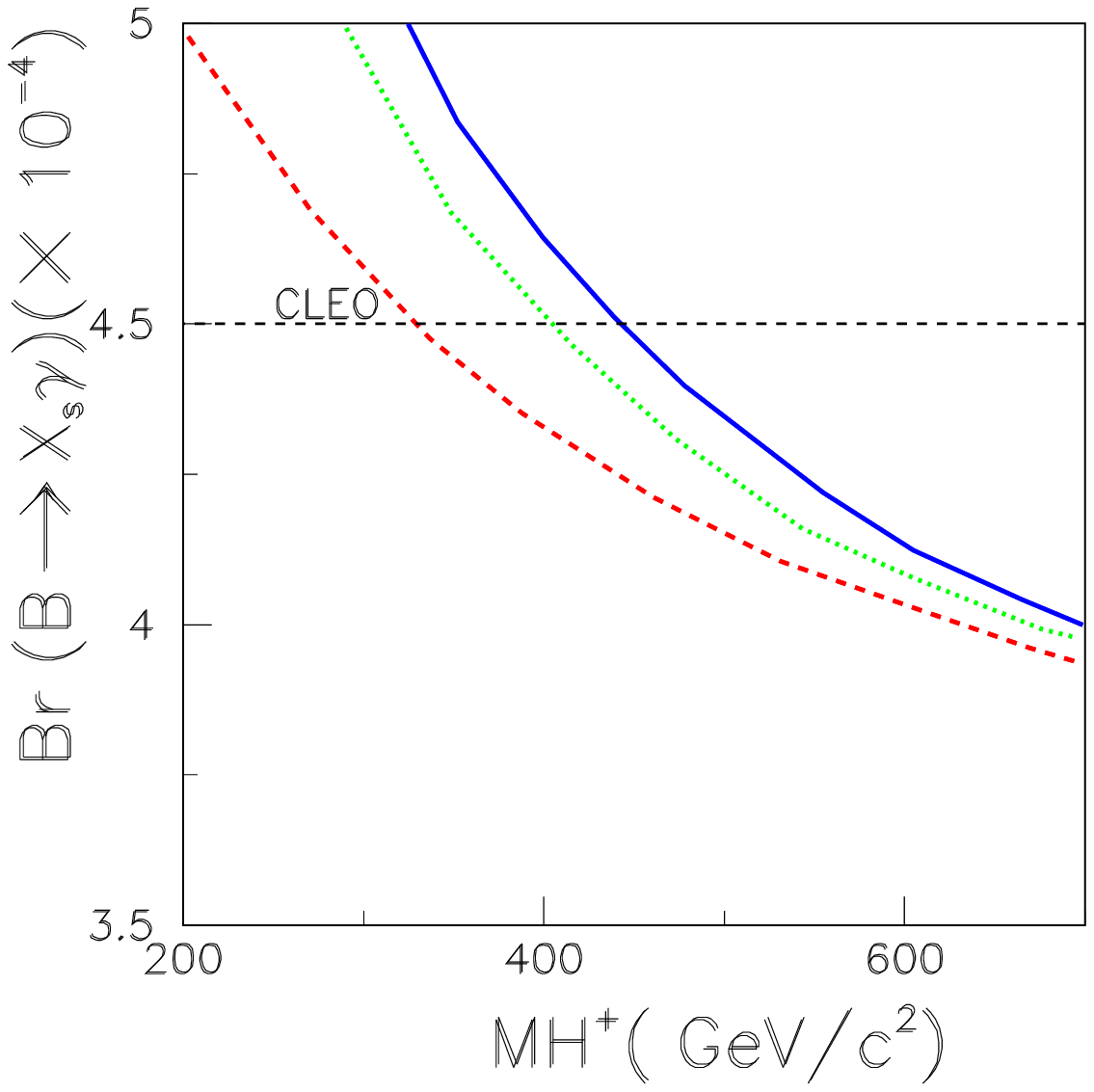,height=2.5cm}}
&
\protect\hbox{\psfig{file=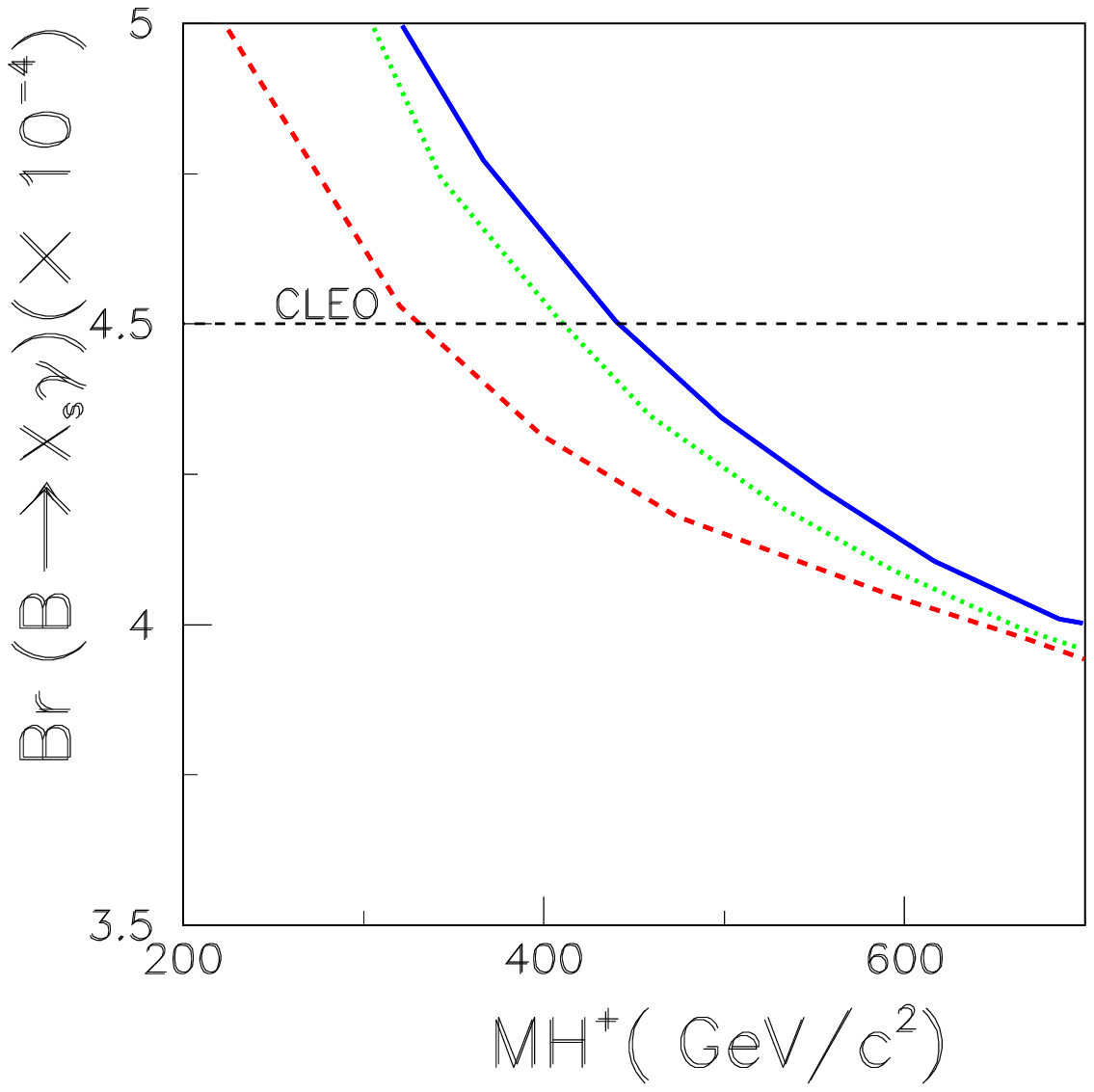,height=2.5cm}}\\
\protect\hbox{\psfig{file=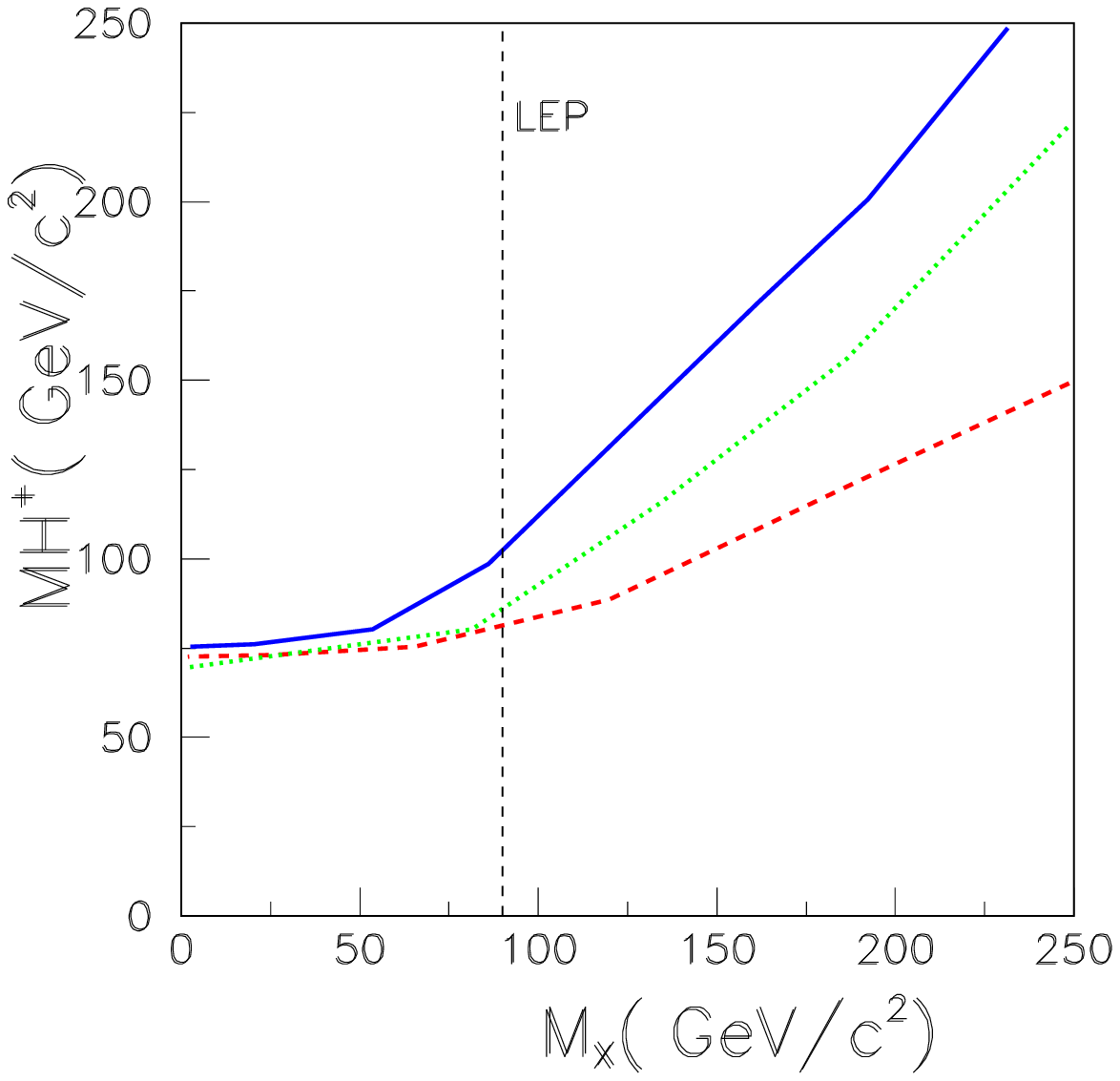,height=2.5cm}}
&
\protect\hbox{\psfig{file=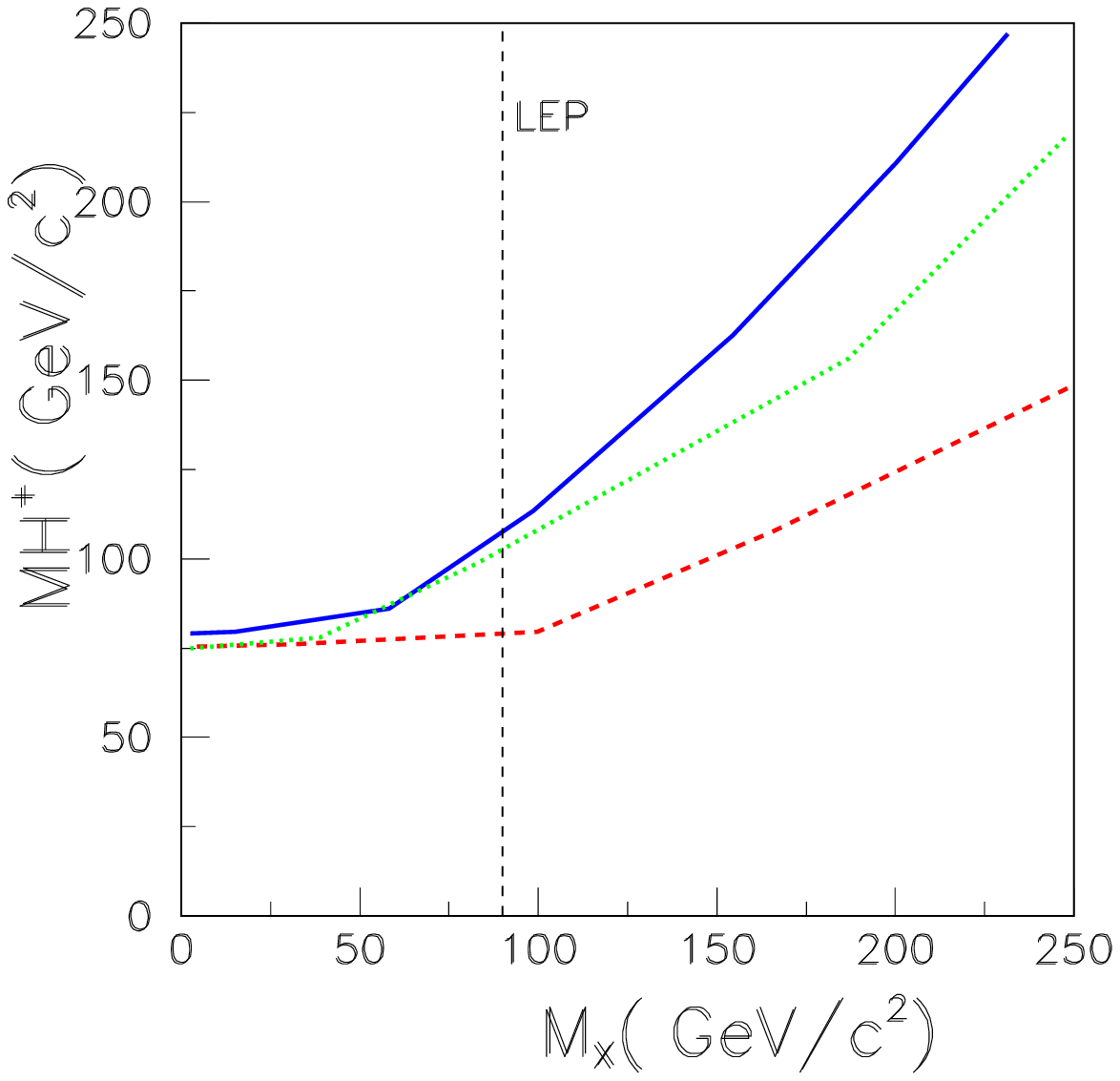,height=2.5cm}} 
\end{tabular}
\caption{
Top and Bottom Left figures: 
Respectively figures \protect\ref{brmh} and \protect\ref{mhstp} for the 
case $m_\nu< 100 $ eV.
Top and Bottom Right figures: 
Respectively figures \protect\ref{brmh} and \protect\ref{mhstp} for the 
case $m_\nu< 100 $ eV and 
``faked'' universality: $\Delta B/\surd \Delta M< 0.05 \surd (1+\tan^2 \beta)$, (see Eqs.(15,16) in Ref.\protect\cite{ferrandis}).
 }
\label{lowneutrino}
}

It is important to take into account the 
 theoretical uncertainties on the calculation of
$B(b \to  s\gamma)$. 
We note that in
implementing the QCD corrections we simply take the $B$ scale $Q_b=5$
GeV (see Ref.~\cite{MPR} for a discussion on the uncertainties of the
QCD corrections to the branching ratio).
If we assume a $10\%$ error, then the bound
on the charged Higgs boson mass in the heavy stop limit within the MSSM
reduces to $m_{H^{\pm}}\gsim 320$ GeV. For the same reason, the 
corresponding bounds on the MSSM--BRpV reduce to 
$m_{H^{\pm}}\gsim 200-250$ GeV, which corresponds to a decrease in
70--120 GeV, \ie, comparable to the values quoted above. No changes are
observed in the case of light charged Higgs limits.

\section{Conclusions}

In summary, we have proved that the bounds on the charged Higgs mass
of the MSSM coming from the experimental measurement of the branching
ratio $B(b \to s\gamma)$ are relaxed if 
we add a single bilinear
R--Parity violating term to the superpotential. 

In the MSSM--BRpV model the staus mix with the charged
Higgs bosons and these contribute importantly to $B(b \to s\gamma)$ in
loops with up--type quarks.  
In an unconstrained
version of the model
 where the
values of all the unknown parameters are free at the weak scale 
we have showed that the bounds on the charged Higgs
boson mass from $B(b\rightarrow s\gamma)$ are relaxed by $\sim 100$ GeV
in the heavy squark limit
(squark masses of a few TeV) where the
chargino contribution is negligible.

Even though in the MSSM--BRpV model the tau lepton
mixes with charginos, implying that the tau-lepton also contributes to
$B(b \to s\gamma)$ in loops with up--type squarks, we have shown that
this contribution is negligible.

In order to have a light
charged Higgs boson in SUSY, its large contribution to $B(b \to
s\gamma)$ can only be compensated by a large contribution from a light
chargino and squark. In order to satisfy the experimental bound on
$B(b \to s\gamma)$ with $m_{\chi^{\pm}_1}>90$ GeV in the MSSM it is
necessary to have $m_{H^{\pm}}\gsim 110$ GeV. In the MSSM--BRpV model
this bound is $m_{H^{\pm}}\gsim 75-85$ GeV, i.e. 25--35 GeV weaker
than in the MSSM.  It is important to note that, in contrast to the
MSSM, charged Higgs boson masses as small as these can be achieved in
MSSM--BRpV already at tree level, as discussed in Ref. \cite{v3cha}.
In this case, charged Higgs lighter that
the $W$--gauge boson are possible and observable at LEP2. 
Nevertheless, R--Parity violating decay modes will compete with
the traditional decay mo\-des of the char\-ged Higgs in the MSSM.

The reason to the relaxation of the MSSM bounds can be understood as
follows: while the charged Higgs couplings to quarks diminish with the
presence of Higgs--Stau mixing, the contribution from the staus not
always compensate this decrease because the stau mass is, in general,
different from the charged Higgs boson mass, and could be larger.

\acknowledgments

I am thankful to my collaborators M.A. Diaz
 and J. Valle for their contribution to the work 
presented here. 
This work was supported by DGICYT grant PB95-1077 and by
the EEC under the TMR contract ERBFMRX-CT96-0090.

\end{document}